%% file: SM0_AHDC_SensorsJ_2011_v9.tex
\documentclass[11pt]{IEEEtran}
\usepackage{graphicx,amssymb,amstext,amsmath,amsfonts,color}

\usepackage{times}
\usepackage[T1]{fontenc}
\newcommand{\calA}{{\mathcal{A}}}
\newcommand{\calY}{{\mathcal{Y}}}
\newcommand{\sE}{{\sf{E}}}                  
\newcommand{\bigtimes}{\mbox{ \Large $\times$}}

\title{Sensor management: Past, Present, and Future}
\author{Alfred O. Hero III, {\em Fellow, IEEE} 
        and Douglas Cochran, {\em Senior Member, IEEE}
\thanks{A.~O.~Hero is with the Department of Electrical Engineering and Computer Science, University of Michigan, Ann Arbor, MI 48109-2122, USA. D.~Cochran is with the School of Mathematical and Statistical Sciences and the School of Electrical, Computer, and Energy Engineering, Arizona State University, Tempe, AZ 85287-5706, USA.}}

\begin{document}

\maketitle

\begin{abstract}
Sensor systems typically operate under resource constraints that
prevent the simultaneous use of all resources all of the time. Sensor
management becomes relevant when the sensing system has the capability
of actively managing these resources; i.e., changing its operating
configuration during deployment in reaction to previous
measurements. Examples of systems in which sensor management is
currently used or is likely to be used in the near future include 
autonomous robots, surveillance and reconnaissance networks, and
waveform-agile radars. This paper provides an overview of the theory,
algorithms, and applications of sensor management as it has developed
over the past decades and as it stands today.
\end{abstract}

\input{SM1_Introduction_v9}
\input{SM2_Description_v9}

\input{SM3_History_v9}

\input{SM4_State_v9}
\input{SM5_Opportunities_v9}

\input{SM6_Conclusion_v9}

 
\bibliographystyle{IEEEtran.bst}
\bibliography{SM7_Refs_v9}

\end{document}

%% file: SM1_Introduction_v9.tex
\section{Introduction}
\label{sec:intro}


Advances in sensor technologies in the last quarter of the 20th century
led to the emergence of large numbers of controllable degrees of freedom
in sensing devices.  Large numbers of traditionally hard-wired 
characteristics, such as
center frequency, bandwidth, beamform, sampling rate, and many other
aspects of sensors' operating modes started to be addressable via
software command. The same period brought remarkable advances in networked
systems as well as deployable autonomous and semi-autonomous vehicles 
instrumented with wide ranges of sensors and interconnected by networks,
leading to configurable networked sensing systems. 
These trends, which affect a broad range of sensor types,
modalities, and application regimes, have continued to the present day and
appear unlikely to abate: new sensing concepts are increasingly 
manifested with device technologies and system architectures
that are well suited to providing
agility in their operation.  

The term ``sensor management,'' as used in this paper, refers to
control of the degrees of freedom in an agile sensor system to satisfy
operational constraints and achieve operational objectives. To accomplish this,
one typically seeks a policy for determining the optimal sensor configuration
at each time, within constraints, as a function of information available from 
prior measurements and possibly other sources. With this perspective, the
paper casts sensor management in terms of formulation and approximation of
optimal planning policies. This point of view has led to a rich vein of
research activity that extends and blends ideas from control, information
theory, statistics, signal processing, and other areas of
mathematical, statistical, and computational sciences and engineering.
Our viewpoint is also slanted toward sensor management in large-scale
surveillance and tracking systems for civilian and defense applications.
The approaches discussed have much broader utility, but the specific
objectives, constraints, sensing modalities, and dynamical models considered
in most of the work summarized here have been drawn from this application
arena. 

Within its scope of attention, the intention of this paper is to
provide a high-level overview; references are given to guide the
reader to derivations of mathematical results, detailed descriptions 
of algorithms, and specifications of application scenarios and systems.
The list of references, while extensive, is not exhaustive; rather it is 
representative of key contributions that have shaped the field and led to
its current state.
Moreover, there are several areas relevant or related to sensor management
that are not within the scope of this survey.  These include purely 
heuristic approaches to sensor management and scheduling as well as
adaptive search methods, clinical treatment planning, 
human-in-the-loop systems such as relevance feedback learning,
robotic vision and autonomous navigation (path planning), 
compressive and distilled sensing, and robust sensing based on
non-adaptive approaches.

The most comprehensive recent survey on sensor management 
of which the authors are aware is the 2008 book \cite{Hero&etal:08}.
This volume consists of chapters written collaboratively by numerous
current contributors to the field specifically to form a perspicuous overview
of the main methods and some noteworthy applications. The 1998 survey paper
by A.~Cassandra \cite{cassandra1998survey}, while not devoted to sensor
management, describes a few applications of partially observed Markov
decision process (POMDP) methods in the general
area of sensor management and scheduling, thereby illustrating conceptual
connections between sensor management and the many other POMDP applications
summarized in the paper. The earlier 1982 survey paper by G.~E.~Monahan 
\cite{Monahan:82} does not consider sensor management applications, but
gives an excellent overview of the base of theory and algorithms for POMDPs
as they were understood a few years before sensor management was becoming 
established as an appreciable area of research.
A 2000 paper by G.~W.~Ng and K.~H.~Ng \cite{ng2000sensor} provides an overview
of sensor management from the perspective of sensor fusion as it stood at
that time. This point of view, although not emphasized in this paper or in 
\cite{Hero&etal:08}, continues to be of interest in the research literature.
Another brief survey from this period is given by X.-X.~Liu {\it et al}.~in 
\cite{Liu:02}, and a short survey of emerging sensor concepts amenable to active sensor
management is given in \cite{Cochran:05}.

Several doctoral dissertations on the topic of sensor management have 
been written in the past fifteen years. Most of these include summaries 
of the state of the art and relevant literature at the time they were 
composed.  
Among these are the dissertations of G.~A.~McIntyre (1998) 
\cite{mcintyre1998comprehensive}, D.~Sinno (2000) \cite{Sinno:00},
C.~M.~Kreucher (2005) \cite{Kreucher:THESIS05}, R.~Rangarajan (2006)
\cite{Rangarajan:THESIS06}, D.~Blatt (2007) \cite{Blatt:THESIS07},
J.~L.~Williams (2007) \cite{williams2007information}, M. Huber
(2009) \cite{huber2009probabilistic}, and K.~L.~Jenkins (2010) \cite{Jenkins:10}. 

The remainder of this paper is organized as follows. 
Section \ref{sec:description} describes the basic goals and defines 
the main components of a sensor management system. In Section 
\ref{sec:history}, the emergence of sensor management is recounted
within a historical context that includes both the advancement
of statistical methods for sequential definition, collection, and
analysis of samples and the rise of sensor technologies and sensing
applications enabling and calling for sensor management. 
Section \ref{sec:soa} gives an overview of some of the
current state of the art and trends in sensor management and
Section \ref{sec:horizon} describes some of the future challenges
and opportunities faced by researchers in the field.

%% file: SM2_Description_v9.tex
\section{Description of sensor management}
\label{sec:description}

The defining function of sensor management is dynamic selection of a sensor, from among a set of available sensors, to use at each time during a measurement period in order to optimize some metric of performance.  Time is usually partitioned into a sequence of epochs and one sensor is to be chosen in each epoch, thereby creating a discrete-time problem.  The term ``sensor management'' most often refers to closed-loop solutions to problems of this nature; i.e, the next sensor to employ is chosen while the sensor system is in operation and in view of the results obtained from prior sensor measurements. The term ``sensor scheduling'' is sometimes used to refer to feed-forward schemes for sensor selection, though this usage is not standardized and the two expressions are used interchangeably in some literature. In current applications of sensor management, and especially in envisioned future applications, the sensors available for selection in each time epoch are actually virtual sensors, each representing one choice of configuration parameters affecting the physical configurations and operating modes of a collection of sensors, sensor suites, sensor platforms, and the way data are processed and communicated among interconnected subsystems.  
With this perspective, selecting a sensor really means determining the values to which the available controllable degrees of freedom in a sensor system should be set.

\begin{figure*}
\centering
\includegraphics[width=6.0in]{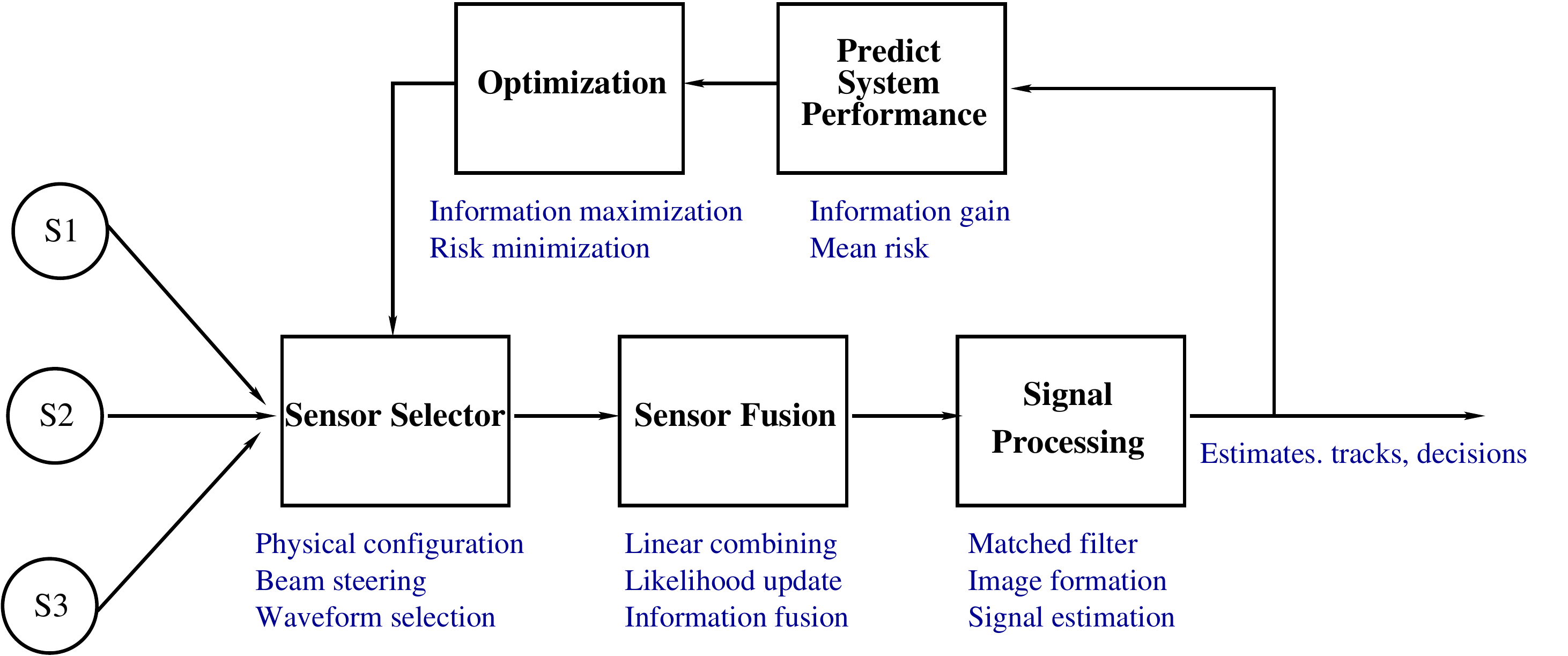}
\caption{
Conceptual block diagram of a sensor management system. The sensor selector selects among sensor actions S1, S2, and S3 based on the output of the optimizer. The optimizer attempts to optimize a system performance metric, such as information gain or mean risk associated with decisions or estimates produced by signal processing algorithms that operate on fused sensor data.}
\label{fig:conceptual_block}
\end{figure*}

Figure \ref{fig:conceptual_block} illustrates the basic elements and operation of a closed-loop sensor management system.  Once a sensor is selected and a measurement is made, information relevant to the sensing objective is distilled from the raw sensor data.  This generally entails fusion of data representing disparate sensing modalities (e.g., optical and acoustic) and other properties, and further combining it with information gleaned from past measurements and possibly also side information from sources extrinsic to the sensor system. The fusion and signal processing components of the loop may produce ancillary information, such as target tracks or decisions about matters external to the sensor manager (e.g., direct an aircraft to take evasive action to avoid collision).  For the purposes of sensor management, they must yield a state of information on the basis of which the merit of each possible sensor selection in the next time epoch may be quantified.  Such quantification takes many forms in current approaches, from statistical (e.g., mean risk or information gain) to purely heuristic. From this point, the sensor manager must optimize its decision as to which sensor to select for the next measurement.

\begin{figure}[b]
\centering
\includegraphics[width=3in]{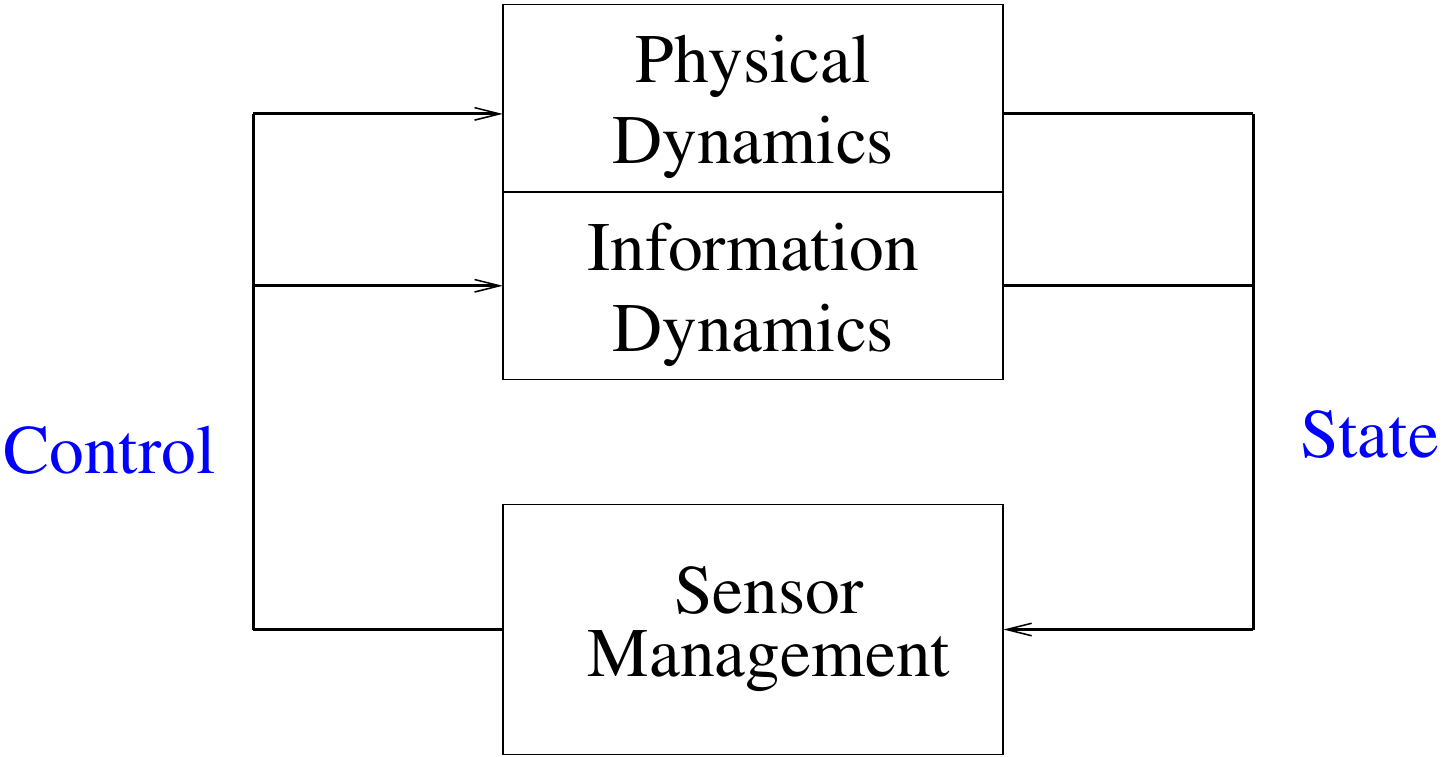}
\caption{A control-theoretic view of sensor management casts the problem as that of optimally controlling a state, sometimes regarded as consisting of separate information and physical components, through the selection of measurement actions.}
\label{fig:state_loop}
\end{figure}

The notion of state is worthy of a few additional words.  Heuristically, the state of information should represent all that is known about the scenario being sensed, or at least all that is relevant to the objective. Often this includes information about the physical state of the sensor system itself (e.g., the position and orientation of the air vehicle carrying one of the video sensors), which may constrain what actions are possible in the next step and thus the set of virtual sensors available to select in the upcoming epoch.  Knowledge of the physical state frequently has utility extrinsic to the sensor manager, so some literature distinguishes physical and information states and their coupled dynamical models as depicted in Figure \ref{fig:state_loop}. This diagram evinces the similarity of sensor management and feedback control in many important respects, and indeed control theory is an important ingredient in current perspectives on sensor management. But sensor management entails certain aspects that give it a distinctive character. Chief among these is in the role of sensing.  In traditional feedback control, sensors are used to ascertain information about the state of a dynamical plant. This information informs the control action through a control law or policy which in turn affects the state.  In sensor management, the state of information is directly affected by the control action; i.e., rather than helping to decide what control action to invoke, the act of sensing is itself the control action.

Sensor management is motivated and enabled by a small number of essential elements. The following paragraphs describe these and explain the roles they play in the current state of the subject. First, a summary of waveform-agile radar is given to provide the context of a current application for the more general descriptions that follow. 

\subsection{Sensor management application -- Waveform-agile radar}

Among the most well developed focus applications of sensor management is real-time closed-loop scheduling of radar resources. The primary feature of radar systems that makes them well suited for sensor management is that they offer several controllable degrees of freedom.  Most modern radars employ antenna arrays for both the transmitter and receiver, which often share the same antenna. This allows the illumination pattern on transmit as well as the beam pattern on receive to be adjusted simply by changing parameters in a combining algorithm.  This ability has been capitalized upon, for example, by adaptive signal processing techniques such as adaptive beamforming on both transmit and receive and more recently by space-time adaptive processing (STAP). The ability for the transmitter to change waveforms in a limited way, such as switching between a few pre-defined waveforms in a library, has existed in a few radar systems for decades. Current radar concepts allow transmission of essentially arbitrary waveforms, with constraints coming principally from hardware limitations such as bandwidth and amplifier power. They also remove traditional restrictions that force the set of transmit antenna elements to be treated as a phased array (i.e., all emitting the same waveform except for phase factors that steer the beam pattern), thereby engendering the possibility of the transmit antennas simultaneously emitting completely different waveforms. This forms the basis of one form of so-called multi-input multi-output (MIMO) radar. 

Two more aspects of the radar application stand out in making it a good candidate for sensor management. One is that pulse-Doppler radars have discrete time epochs intrinsically defined by their pulse repetition intervals and often also by their revisit intervals \cite{Richards:05}. Also, in radar target tracking applications there are usually well defined performance metrics and well developed dynamical models for the evolution of the targets' positions, velocities, and other state variables.  These metrics and models directly enhance the sensor manager's ability to quantitatively predict the value of candidate measurements before they are taken. 

In view of these appealing features, it is no surprise that radar applications have received a large amount of attention as sensor management has developed.
The idea of changing the transmitted waveform in a radar system in an automated fashion in consideration of the echo returns from previously transmitted waveforms dates to at least the 1960s, though most evidence of this is anecdotal rather than being documented in the research literature. The current generation of literature on closed-loop waveform management as a sensor management application began with papers of D.~J.~Kershaw and R.~J.~Evans \cite{Kershaw:94,Kershaw:97} and S.~M.~Sowelam and A.~H.~Tewfik \cite{Sowelam:94,Sowelam:97} in the mid-1990s, roughly corresponding to the ascension of sensor management literature in broader contexts. Among the early sensor management papers that focused on closed-loop beam pattern management were those of V.~Krishnamurthy and Evans in the early 2000s \cite{krishnamurthy2001hidden,krishnamurthy2003correction}.
Several contributions by numerous authors on these and related radar sensor management applications have appeared in the past decade. 
Among the topics addressed in this recent literature are radar waveform scheduling 
for target identification \cite{Goodman:07}, target tracking \cite{sira:09spm}, clutter and interference mitigation \cite{Greco:08,Sira:07}, and simultaneously estimating and tracking parameters associated with multiple extended targets \cite{Leshem:07}. There has also been recent interest in drawing insights for active radar and sonar sensor management from biological echolocation systems \cite{Vespe:09} and in designing optimal libraries of waveforms for use with radar systems that support closed-loop waveform scheduling \cite{cochran2009waveform}.

\subsection{Controllable Degrees of Freedom}

Degrees of freedom in a sensor system over which control can be exercised with the system in operation provide the mechanism through which sensors can be managed.
In envisioned applications, they include diverse sets of parameters, including physical configuration of the sensor suite, signal transmission characteristics such as waveform or modulation type, signal reception descriptors ranging from simple on/off state to sophisticated properties like beamform. 
They also include algorithmic parameters that affect local versus centralized processing trade-offs, data sharing protocols and communication schemes, and typically numerous signal processing choices.

Many characteristics of current and anticipated sensor systems that are controllable during real-time operation were traditionally associated with subsystems that were designed independently.  Until relatively recently, transduction of physical phenomena into electrical signals, analog processing, conversion to digital format, and digital processing at various levels of information abstraction were optimized according to  performance criteria that were often only loosely connected with the performance of the integrated system in its intended function. Further integrated operation of such subsystems generally consisted of passing data downstream from one to the next in a feed-forward fashion. Integrated real-time authority over controllable degrees of freedom spanning all of this functionality not only allows joint optimization of systemic performance metrics but also accommodates adaptation to changing objectives.

In the radar sensor management example, the ease and immediacy of access (i.e., via software command) to crucial operating parameters such as antenna patterns and waveforms provides the means by which a well conceived algorithm can manage the radar in each time epoch.

\subsection{Constraints}

The utility of sensor management emerges when it is not possible to process, or even collect, all the data all the time. Operating configurations of individual sensors or entire sensor systems may be intrinsically mutually exclusive; 
e.g., the transmitter platform can be in position $A$ or in position $B$ at the time the next waveform is emitted, but not both. 
One point of view on configurable sensors, discussed in \cite{Cochran:98}, imagines an immense suite of virtual sensor systems, each defined by a particular operating configuration of the set of physical sensors that comprises the suite. Limitations preventing an individual sensor from being in multiple configurations at the same time are seen as constraints to be respected in optimizing the configuration of the virtual sensor suite. This is exactly the case in the waveform-agile radar example, where only one waveform can be transmitted on each antenna element at any given time.

Restrictions on communications and processing resources almost always constrain what signal processing is possible in networked sensor applications. Collecting all raw data at a single fusion center is seldom possible due to bandwidth limitations, and often to constraints imposed by the life and current production of batteries as well. So it is desirable to compress raw data before transmission.  But reducing the data at the nodes requires on-board processing, which is typically also a limited resource.

\subsection{Objective Quantification}

When controllable degrees of freedom and constraints are present, sensor management is possible and warranted. In such a situation, one would hope to treat the selection of which sensing action to invoke as an optimization problem. But doing so requires the merit of each possible selection to be represented in such a way that comparison is possible; e.g., by the value of a cost or objective functional.  

The value of a specified set of data collection and processing choices generally depends on what is to be achieved.  
For example, one set of measurements by a configurable chemical sensor suite may be of great value in determining whether or not an analyte is an explosive, but the best data to collect to determine the species of a specimen already known to be an explosive may be quite different. 
Moreover, the objective may vary with time or state of knowledge: once a substance is determined to be an explosive, the goal shifts to determining what kind of explosive it is, then how much is present, then precisely where it is located, etc. Consequently, predictively quantifying the value of the information that will be obtained by the selection or a particular sensing action is usually difficult and, at least in principle, requires a separate metric for each sensing objective that the system may be used to address. The use of surrogate metrics, such as information gain discussed in Section \ref{sec:soa}, has proven effective in some applications. With this approach, the role of a metric designed specifically for a particular sensing objective is undertaken by a proxy, usually based on information theoretic measures, that is suited to a broader class of objectives. This approach sacrifices specificity in exchange for relative simplicity and robustness, especially to model mismatch.

Management of radar beamforms and waveforms for target tracking, though not trivial, is one of the most tractable settings for objective quantification.  The parameters can be chosen to optimize some function of the track error covariance, such as its expected trace or determinant, at one or more future times; e.g., after the next measurement, after five measurement epochs, or averaged over the next ten epochs. Computation or approximation of such functions is assisted by the tracker's underlying model for the dynamical evolution of the target states. The use and effectiveness of waveform management in such applications is discussed in \cite[Ch.~10]{Hero&etal:08}, which also cites numerous references.



%% file: SM3_History_v9.tex
\section{Historical roots of sensor management}
\label{sec:history}

It has long been recognized that appropriate collection of data is essential in the design of experiments to test hypotheses and estimate quantities of interest. R.~A.~Fisher's classical work \cite{Fisher:35}, which encapsulated most of the ideas on statistical design of experiments developed through the first part of the 20th century, primarily addressed the situation in which the composition of the sample to be collected is to be determined in advance of the experiment.  In the early 1950s, the idea of using closed-loop strategies in experiment design emerged in connection with sequential design of experiments.  In his 1951 address to the Meeting of the American Mathematical Society \cite{Robbins:52}, H.~Robbins observed:
\begin{quote}
A major advance now appears to be in the making with the creation of a theory of the
sequential design of experiments, in which the size and composition of the samples are not fixed in advance but are functions of the observations themselves.
\end{quote}
Robbins attributes the first application of this idea to Dodge and Romig in 1929 \cite{Dodge:29} in the context of industrial quality control.  They proposed a double sampling scheme in which an initial sample is collected and analyzed, then a determination about whether to collect a second sample is based on analysis of the first sample. This insight was an early precursor to the development of sequential analysis by Wald and others during the 1940s \cite{Wald:47}, and ultimately to modern methods in statistical signal processing such as sequential detection \cite{Helstrom:94}.
In the interim, H.~Chernoff made substantial advances in the statistical study of optimal design of sequences of experiments, particularly for hypothesis testing and parameter estimation \cite{Chernoff:53,Chernoff:59}. Many results in this vein are included in his 1972 book \cite{Chernoff:72}. Also in 1972, V.~V.~Fedorov's book \cite{Fedorov:72} presented an overview of key results, many from his own research, in optimal experimental design up to that time.  The relevance of a portion of Fedorov's work to the current state of sensor management is noted in Section \ref{sec:soa}.

One view of the {\em raison d'\^etre} for sensors, particularly among practitioners of sensor signal processing, is to collect samples to which statistical tests and estimators may be applied. From this perspective, the advancement of sensor signal processing over the latter half of the 20th century paralleled that of experimental design.  By the early 1990s, a rich literature on detection, estimation, classification, target tracking and related problems had been compiled.  Nearly all of this work was predicated on the assumption that the data were given and the goal was to process it in ways that are optimally informative in the context of a given application.  There were a few notable cases in which it was assumed the process of data collection could be affected in a closed-loop fashion based on data already collected.  In sequential detection theory, for example, the data collection is continued or terminated at a given time instant (i.e., binary feedback) depending on whether a desired level of confidence about the fidelity of the detection decision is supported by data already collected. An early example of closed-loop data collection involving a dynamic state was the ``measurement adaptive problem'' treated by L.~Meier {\it et al.}~in 1967 \cite{Meier:67}. This work sought to simultaneously optimize control of a dynamic plant and the process of collecting  measurements for use in feedback. 
{
Another is given in a 1972 paper of M.~Athans \cite{Athans:72} that considers optimal closed-loop selection of the linear measurement map in a Kalman filtering problem.

One of the first contexts in which the term ``sensor management'' was used in the sense of this discussion\footnote{The phrase comes up in various literature in ways that are related to varying degrees to our use in this paper. To maintain focus, we have omitted loosely related uses of the term, such as in clinical patient screening applications \cite{Landoll:69}.} was in automating control of the sensor systems in military aircraft (see, e.g., \cite{Bier:1988}).  In this application, the constrained resource is the attention of the pilot, particularly during hostile engagement with multiple adversaries, and the objective of sensor management is to control sensor resources in such a way that the most important information (e.g., the most urgent threats) are emphasized in presentation to the pilot. Applications associated with situational awareness for military aircraft continue to be of interest, and this early vein of application impetus expanded throughout the 1990s to include scheduling and management of aircraft-based sensor assets for surveillance and reconnaissance missions (see, e.g., \cite{musick1994chasing,malhotra1997learning} and \cite[ch.~11]{Hero&etal:08}).

Also beginning in the 1980s, sensor management was actively pursued under the label of ``active vision'' for applications in robotics \cite{Aloimonos:1988}. This work sought to exercise feedback control over camera direction and sometimes other basic parameters (e.g., zoom or focal distance) to improve the ability of robotic vision systems to contribute to navigation, manipulation, and other tasks entailed in the robot's intended functionality.

The rapid growth of interest in sensor management beginning in the 1990s can be attributed in large part to developments in sensor and communications technologies. New generations of sensors, encompassing numerous sensing modalities, are increasingly agile. Key operating parameters, once hard-wired, can be almost instantly changed by software command. Further, transducers can be packaged with A/D converters and microprocessors in energy efficient configurations, in some cases on a single chip, creating sensors that permit on-board adaptive processing involving dynamic orchestration of all these components. At the same time, the growth of networks of sensors and mobile sensor platforms is contributing even more controllable degrees of freedom that can be managed across entire sensor systems.  From a purely mathematical point of view, it is almost always advantageous to collect all available data in one location (i.e., a ``fusion center'') for signal processing. In today's sensor systems, this is seldom possible because of constraints on computational resources, communication bandwidth, energy, deployment pattern, platform motion, and many other aspects of the system configuration. Even highly agile sensor devices are constrained to choose only one configuration from among a large collection of possibilities at any given time. 

These years spawned sensor management approaches based on the modeling sensor management as a decision process, a perspective that underpins most current methods as noted in Section \ref{sec:soa}. Viewing sensor management in this way enabled tapping into a corpus of knowledge on control of decision processes, Markov decision processes in particular, that was already well established at the time \cite{Segall:77}. Initial treatments of sensor management problems via POMDPs, beginning with D.~Casta\~n\'on's 1997 paper \cite{castanon1997approximate}, were followed shortly by other POMDP-based ideas such as the work of J.~S.~Evans and Krishnamurthy published in 2001--2002 \cite{Evans:01,krishnamurthy2002algorithms}. These were the early constituents of a steady stream of contributions to the state of the art summarized in Section \ref{sec:MDP}. A formidable obstacle to the practicality of the POMDP approach is the computational complexity entailed in its implementation, particularly for methods that look more than one step ahead. Consequently, the need for approximation schemes and the potential merit of heuristics to provide computational tractability was recognized from the earliest work in this vein.

The multi-armed bandit (MAB) problem is an important exemplar of a class of multi-stage decision problems where actions yielding large immediate rewards must be balanced with others whose immediate rewards are smaller, but which hold the potential for greater long-term payoff. 
While two-armed and MAB problems had been studied in previous literature, the origin of index policy solutions to MAB problems dates to J.~C.~Gittins in 1979 \cite{Gittins:79}.
 As discussed in Section \ref{sec:MAB}, under certain assumptions, an index solution assigns a numerical index to each possible action at the current stage of an infinitely long sequence of plays of a MAB.  The indices can be computed by solving a set of simpler one-armed bandit problems and their availability reduces the decision at each stage to choosing the action with the largest index. The optimality of Gittins' index scheme was addressed by P. Whittle in 1980 \cite{Whittle:80}.

As with POMDPs, the MAB perspective on sensor management started receiving considerable research attention around 2000. Early applications of MAB methodology to sensor management include the work of Krishnamurthy and R.~J.~Evans \cite{krishnamurthy2001hidden,krishnamurthy2003correction} who considered a multi-armed bandit model with Markov dynamics for radar beam scheduling. The 2002 work of R.~Washburn {\it et al.}~\cite{washburn2002stochastic}, although written in the context of more general dynamic resource management problems, was influential in the early develop of MAB approaches to sensor management.

A theory of information based on entropy concepts was introduced by C.~E.~Shannon in his classic 1948 paper \cite{Shannon:48} and was subsequently extended and applied by many others, mostly in connection with communication engineering. Although Shannon's theory is quite different than that of Fisher, sensor management has leveraged both in various developments of information-optimized methods. These were introduced specifically to sensor management in the early 1990s by J.~Manyika and H.~Durrant-Whyte \cite{manyika1995data} and by W.~W.~Schmaedeke \cite{schmaedeke1993information}. As remarked in Section \ref{sec:soa}, information-based ideas were applied to particular problems related to sensor management even earlier. Fisher's information theory was instrumental in the development of the theory of optimal design of experiments, and numerous examples of applications of this methodology have appeared since 2000; e.g., \cite{zhang2004detection,tian2006sensor}. Measures of information led to sensor management schemes based on information gain, which developed into one of the central thrusts of sensor management research over the past decade. Some of this work is summarized in Section \ref{sec:soa_information}, and a more complete overview of these methods in provided in \cite[Ch.~3]{Hero&etal:08}.

From foundations drawing on several more classical fields of study, sensor management has developed into a well-defined area of research that stands today at the crossroads of the disciplines upon which it has been built. Key approaches that are generally known to researchers in the area are discussed in the following section of this paper. But sensor management is an active discipline, with new work and new ideas appearing regularly in the literature. Some noteworthy recent developments include work by V.~Gupta {\it et al.}~which introduces random scheduling algorithms that seek optimal mean steady state performance in the presence of probabilistically modeled effects \cite{Gupta:06,Gupta:04,Chung:04}. K.~L.~Jenkins {\it et al.}~very recently proposed the use of random set ideas, similar to those applied in some approaches to multi-target tracking, in sensor management \cite{Jenkins:ACC10,Jenkins:CDC10}. These preliminary investigations have resulted in highly efficient algorithms for certain object classification problems. Also very recently, D.~Hitchings {\it et al.}~introduced new stochastic control approximation schemes to obtain tractable algorithms for sensor management based on receding horizon control formulations \cite{Hitchings:10}.  They also proposed a stochastic control approach for sensor management problems with large, continuous-valued state and decision spaces \cite{Hitchings:CDC10}. 

Despite ongoing progress, sensor management still holds many unresolved challenges.
Some of these are discussed in Section \ref{sec:horizon}.

%% file: SM4_State_v9.tex
\section{State of the art in sensor management}
\label{sec:soa}

The theory of decision processes provides a unifying perspective for
the state of the art in sensor management research today.  A decision
process, described in more detail below, is a time sequence of
measurements and control actions in which each action in the sequence
is followed by a measurement acquired as a result of the previous
action.  With this perspective, the design of a sensor manager is
formulated as the specification of a decision rule, often called a policy,
that generates realizations of the decision process. An optimal policy
will generate decision processes that, on the average, will maximize
an expected reward; e.g., the negative mean-squared tracking error or
the probability of detection.  A sound approach to
sensor management will either approximate an optimal policy in some
way or else attempt to analyze the performance of a proposed heuristic
policy. In this section we will describe some current approaches to
design of sensor management policies. The starting point is
a formal definition of a decision process.

\subsection{Sensor management as a decision process}
\label{sec:DP}

Assume that a sensor collects a data sample $y_{t+1}$ at time $t$
after taking a sensing action $a_t$. It is typically assumed that the
possible actions are selected from a finite action space
$\calA$, that may change over time. The selected action $a_k$ depends
only on past samples $\{y_k, y_{k-1}. \ldots, y_1\}$ and past actions
$\{a_{k-1}, a_{k-2}. \ldots, a_0\}$, and the initial action $a_0$ is
determined offline. The function that maps previous data samples and
actions to current actions is called a policy.  That is, at any time
$t$, a policy specifies a mapping $\gamma_t$ and, for a specific set of
samples, an action $a_t=\gamma_t(\{a_k\}_{k<t}, \{y_k\}_{k\leq t})$.
A decision process is a sequence $\{a_k, y_{k+1}\}_{k\geq 0}= \{a_0,
y_1, a_1, y_2, a_2, y_3\ldots, \}$, which is typically random and can
be viewed as a realization from some generative model specified by the
policy and the sensor measurement statistics.

A well designed sensor manager will formulate the  policy with the
objective of maximizing an average reward. The reward at time $t$ is a
function $R_t(\{a_k\}_{k<t}, \{s_k\}_{k\leq t})$ of the action
sequence $\{a_k\}_{k>0}$ and a state sequence $\{s_k\}_{k>0}$,
describing the environment or a target in the environment.  The state
$s_k$ might be continuous (e.g., the position of a moving target) or
discrete (e.g., $s_k=1$ when the target is moving and $s_k=0$ when it
is not moving).  It is customary to model the state as random and the
data sample $y_k$ as having been generated by the state $s_k$ in some random
manner. In this case, there exists a conditional distribution of the
state sequence given the data sequence and the average reward at time
$t$ can be defined through the statistical expectation
$\sE[R_t(\{a_k\}_{k<t}, \{s_k\}_{k\leq t})]$.

An optimal action policy will maximize the average award at each time
$t$ during the sensor deployment time period. The associated
optimization must be performed over the set of mappings $\gamma_t$
defined on the cartesian product spaces $\bigtimes_{k=1}^{t}
\left\{\calA_{k-1} \bigtimes \calY  \right\}$ and mapping to $\calA_t$ for $t=0, 1, \ldots$.
Due to the high dimensionality of the cartesian product spaces, no tractable
methods exist for determining optimal action policies under this degree
of generality.  Additional assumptions on the statistical
distributions of the decision process and state process are needed to
reduce the dimensionality of the optimization spaces.

When the unknown state $s_k$ is not recoverable from $y_k$ then the
decision process is called a partially observable decision
process. The partially observable case is common in actual sensing
systems where the measurements $y_k$ are typically contaminated by
noise or clutter. However, policy
optimization generally presents more mathematical difficulties in the
partially observable case than in the perfectly observable case.

\subsection{Markov decision processes}
\label{sec:MDP}

A natural way to simplify the task of policy optimization is to assume
that the general decision process described in Section \ref{sec:DP}
satisfies some additional Markovian properties. To make the general
decision process Markovian one imposes the assumption that the state
sequence is dependent only on the most recent state and action given
the entire past. Specifically, we assume that $P(s_{t+1}|\{s_k,
a_k\}_{k\leq t})= P(s_{t+1}|s_t, a_t)$, the conditional state
transition probability, and $P(y_t|\{s_k, a_k\}_{k\leq t})= P(y_t|s_t,
a_t)$, the measurement likelihood function given action $a_t$.

We additionally restrict the reward to be additive over time and only
consider policies that depend on the most recent measurement, i.e.,
$R_t(\{a_k\}_{k<t}, \{s_k\}_{k\leq t})=\sum_{k=0}^t R_t(a_k, s_k) $
and the associated mapping $\gamma_t$ is restricted to be from
$\calY\bigtimes \calA_{t-1}$ to $\calA_t$.  When the state can be
recovered from the measurements the resultant process is called a
Markov decision process (MDP). When the state is not recoverable from
the measurements the resultant process is called a partially
observable Markov decision process (POMDP).

For MDP or POMDP models the optimal restricted policy can be
determined by backwards induction over time. In particular, there is a
compact recursive formula, known as Bellman's equation, for determining
the mapping $\gamma_{t-1}$ from the mapping $\gamma_{t}$.  In
special cases where the state and the measurements obey standard
dynamical stochastic state models (e.g., the linear-Gaussian
model assumed in the
Kalman filter), this optimal restricted policy is in fact the overall
optimal policy.  That is, the overall optimal policy only depends on
the most recent measurements.  Furthermore, as shown by E.~J.~Sondik
\cite{Sondik71}, the optimal policy can be found by linear
programming. For more details on MDPs, POMDPs, and Bellman's equation
and solutions, the reader is referred to \cite[Ch.~2]{Hero&etal:08}.

As noted in Section \ref{sec:history},
the application of MDP and POMDP methods to sensor management problems
can be traced back to the mid 1990's.  
{In their 1994 overview of the
field of sensor management \cite{musick1994chasing}, S.~Musick and R.~Malhotra suggested
that a comprehensive mathematical framework was needed to assess and
optimize scheduling over sensor and inter-sensor actions.
Anticipating the future application of POMDP and reinforcement learning approaches,
went on to suggest adaptive control, state space
representations, and mathematical programming as the components of a
promising framework. However, to be applied to practical large scale
sensor management problems approximate solutions to the POMDP would be
necessary.  Casta\~{n}\'{o}n's 1997 policy rollout approximation
\cite{castanon1997approximate} was the earliest successful application
of the POMDP to sensor management.

Several types of approximations to the optimal POMDP sensor management
solution are discussed in \cite[Ch.~2]{Hero&etal:08} under the heading
of approximate dynamic programming. These include: offline learning,
rollout, and problem approximation techniques. Offline learning
techniques use offline simulation to explore the space of policies and
include the large class of reinforcement learning methods
\cite{malhotra1997learning,Kaelbling98,chong2009des}.
Rollout uses real-time simulation to approximate the rewards of a
sub-optimal policy
\cite{castanon1997approximate,li2009approximate}. Problem
approximation uses a simpler approximate model or reward function for
the POMDP as a proxy for the original problem and includes bandit and
information gain approaches, discussed in the following subsections.


POMDP approaches have been applied to many different sensing systems.
One of the most active areas of application has been distributed
multiple target tracking, see for example \cite{liu2007multitarget}
and references therein. When target dynamics are non-linear and
environments are dynamically changing, the states of targets can be
tracked by a particle filter \cite{Kreucher&etal:AES05}. This filter
produces an estimate of the posterior density of the target tracks
that is used by the scheduler to predict the value of different
sensing actions \cite{chong2009des}.  
Managers for many other sensing
systems have been implemented using POMDPs and reinforcement learning, for
example, multifunction radar \cite{krishnamurthy2009optimal},
underwater sensing applications \cite{ji2007nonmyopic}, passive radar
\cite{hanselmann2008sensor}, and air traffic management
\cite{kochenderfer2008comprehensive}.

\subsection{Multi-armed bandit decision processes}
\label{sec:MAB}

A multi-armed bandit (MAB) is a model for sequential resource
allocation in which multiple resources (the arms of the bandit) are
allocated to multiple tasks by a controller (also called a processor).
When a particular arm $a_t$ of the bandit is pulled (a control action
called a ``play'') at time $t$ the MAB transitions to a random state
$x_t$ and pays out a reward depending on the state. As in a MDP,
successive MAB control actions produce a sequence of actions and
states. When the MAB action-state sequence is Markovian it is a
special case of a MDP or POMDP process.

In some cases, the optimal policy for a $k$-arm MAB problem can be
shown to reduce to a so-called index policy. An index policy is a
simpler mapping that assigns a score (or index) to each arm of the MAB
and pulls only the arm having maximum score at a given time. The key
to the simplification is that these scores, the Gittins indices
mentioned in Section \ref{sec:history},
can be determined by solving a much simpler set of $k$ different
single-armed bandit problems. Gittins index policies exist when the
actions are not irrevocable; meaning that any available actions not
taken at the present time can be deferred to the future, producing the same
sequence of future rewards, except for a discount factor. The
significance of Gittins index policies is that they are frequently
much simpler to compute than backwards induction solutions to optimal
policies for MDPs and POMDPs. Thus they are sometimes used to
approximate these optimal policies; e.g., using rollout with MAB
index-rules as the base policy \cite{Bertsekas&Castanon:Heuristics99}.
See \cite[Ch.~6]{Hero&etal:08} for further discussion of index
policies and their variants.

As a simple example, consider the aforementioned wide area search
problem for the case of a single non-moving target that could be
located in one of $k$ locations with equal probability. Assume that in
each time epoch a sensor can look at a single location with specified
probabilities of correct detection and false alarm.
Further assume that the reward is decreasing
in the amount of time required by the sensor to correctly find the
target. Identify each sensing action (location) as an arm of the MAB
and the un-normalized posterior probability of the true target location
as the state of the MAB.  Under these assumptions, the optimal MAB
policy for selecting arms is an index policy and specifies the optimal
wide area search scheduler.  For further details on this application
of MAB to sensor management see \cite[Ch.~7]{Hero&etal:08}.

Bandit models were proposed for search problems like the above several
decades ago \cite{benkoski1991survey}, but their application to sensor
management is relatively recent.  Early applications of the
multi-armed bandit model to sensor management were Krishnamurthy's
treatment of the radar beam scheduling for multiple target tracking
problem \cite{krishnamurthy2001hidden,krishnamurthy2003correction}
and Washburn {\it et al.}'s
application to general problems of sensor resource management
\cite{washburn2002stochastic}.  As another example, arm acquiring
bandits have been proposed by Washburn \cite[Ch.~7]{Hero&etal:08} for
tracking targets that can appear or disappear from the scene. Also
discussed in \cite[Ch.~7]{Hero&etal:08} are restless bandits, multi-armed
bandits in which the states of the arms not played can evolve in time.
Sensor management application of restless bandits include radar sensor management
for multi-target tracking (see, e.g., \cite{la2006optimal}).

\subsection{Information-optimized decision processes}
\label{sec:soa_information}

The MDP/POMDP and MAB approaches to sensor management involve
searching over multi-stage look-ahead policies. Designing a
multi-stage policy requires evaluating each available action in terms
of its impact on the potential rewards for all future actions. Myopic
sensor management policies have been investigated as low complexity
alternatives to multi-stage policies. Myopic policies only look ahead
to the next stage; i.e., they compute the expected reward in the
immediate future to determine the best current action. Such greedy policies
benefit from computational simplicity, but at the expense of
performance loss compared to multi-stage optimal policies. Often
this loss is significant. However, there are cases where the myopic
loss approach gives acceptable performance, and indeed is almost optimal
in special cases.

The most obvious way to obtain myopic sensor scheduling policies is to only
consider the effect of the control action on the immediate reward;
i.e., to truncate the future reward sequence in the multi-stage POMDP
scheduling problem. This approach is called the optimal one-step
look-ahead policy. However, it has often been observed that a myopic
policy can achieve better overall performance by maximizing a surrogate reward,
such as the mutual information between the data and the target. The
information gain, discussed in more detail below, has the advantage
that it is a more fundamental quantity than a task-specific reward
function. For example, unlike many reward functions associated with
estimation or detection algorithms, the mutual information is
invariant to invertible transformations of the data. This and other
properties lead to myopic policies that are more robust to factors
such as model mismatch and dynamically changing system objectives
(e.g., detection versus tracking), while ensuring a minimal level of
system performance. For further motivation and properties of
information theoretic measures for sensor management the reader may
wish to consult \cite[Ch.~3]{Hero&etal:08}.

Information theoretic measures have a long history in sensor
management. Optimization of Fisher information was applied to the
related problem of optimal design of experiments (DOE) by Fisher,
discussed in Section \ref{sec:history},
in the early part of the twentieth century \cite{Fisher:35}. Various
functions of the Fisher information matrix, including its determinant
and trace, have been used as reward functions for optimal DOE
\cite{Fedorov:72}. More recently, sensor management applications of
optimal DOE have been proposed; e.g., in managed sensor fusion
\cite{manyika1828information}, in sensor managed unexploded
ordnance (UXO) detection
\cite{zhang2004detection}, in multi-sensor scheduling
\cite{hernandez2004multisensor}, and in sensor management for robotic
vision and navigation \cite{tian2006sensor}. However, Fisher
information approaches to sensor management have several drawbacks.  Notable
among these are that the Fisher information requires specification of a parametric model
for the observations. It is also a local measure of
information that does not apply to discrete targets or mixtures of
discrete and continuous valued targets. Model mismatch and/or discrete
valued quantities frequently arise in sensor management
applications. For example, discrete values arise when there is
categorical side information about the target or clutter, or a target
that transitions between two states like stopping and moving. These
are principal reasons that non-local information measures such as entropy and
mutual information have become more common in sensor management.

In his 1998 PhD thesis \cite{mcintyre1998comprehensive}, McIntyre cites
the work of
Barker \cite{barker1977information} and Hintz and McVey
\cite{HintzMcvey91} as the first to apply entropy to sensor management
problems in 1977 and 1991, respectively. However, while the
problems they treated are special cases of sensor management, they did
not treat the general sensor management problem nor did they use the
term in their papers.  The first papers we know of that applied entropy measures
explicitly to sensor management were Manyika and Durrant-Whyte
\cite{manyika1828information} in 1992 and Schmaedeke
\cite{schmaedeke1993information} in 1993. The information measure used
in these papers was the expected update in posterior entropy, called
the information gain, that is associated with a given candidate sensor
action.

These early information theoretic sensor management papers assumed
Gaussian observations and linear dynamics, in which case the entropy
and information gain have closed form mathematical expressions.
Subsequently, the linear-Gaussian assumptions have been relaxed by
using non-parametric estimation of entropy and information
gain. Other information gain measures have also been introduced, such
as the Kullback-Leibler (KL) divergence, the KL discrimination, and
the R\'enyi entropy. The reader can consult the book
\cite{Hero&etal:08} and, in particular, early papers by Schmaedeke and
K.~Kastella
\cite{Schmaedeke&Kastella:SPIE94,kastella1997discrimination},
R.~Mahler \cite{Mahler:SF96}, Hintz and McIntyre
\cite{hintz1998information}, and Kreucher {\it et al.}
\cite{Kreucher&etal:SPIE03_2,Kreucher&etal:IPSN03}.

At first information gain sensor management methods were focused on
single modality tracking of simple passive targets. In recent years,
information gain has been applied to increasingly general models and
sensor management tasks. For example, information driven methods have
been applied to dynamic collaborative sensing with communication costs
\cite{zhao2002information}, multi-sensor information fusion
\cite{xiong2002multi}, target tracking with uncertain sensor responses
\cite{kolba2007information}, multi-target tracking in large dynamic
sensor networks \cite{Kreucher&etal:IEEEProc07}, multi-modality
multi-target tracking with time varying attenuation and obscuration
\cite{kreucher2005smu,chong2009des}, robot path planning
\cite{zhang2009information}, and active camera control for object
recognition and tracking using mutual information
\cite{denzler2002information}.

A striking mathematical result on the capabilities of information
driven sensor management was obtained by J.~L.~Williams \textit{et al.}
\cite{Williams_Fisher:07AISTATS} in connection with the general problem of
information gathering in the context of graphical models under the
assumption of conditionally independent measurements.   
In 2005
Guestrin {\it et al.}~\cite{Guestrin:05} showed that the conditional
mutual information is submodular in the context of general machine learning problems. 
The significance of this result for sensor management is that the maximizer of a submodular objective function can be well approximated using greedy optimization algorithms. 
Using this insight, Williams \textit{et al.}
established in \cite{Williams_Fisher:07AISTATS}
that greedy sequential methods for measurement planning are guaranteed to
perform within a factor of $1/2$ of the optimal multi-stage selection
method.
Furthermore, this bound is independent of the length
of the planning horizon and is sharp. 
The remarkable results of \cite{Williams_Fisher:07AISTATS} 
are significant in that they provide theoretical
justification for the computationally simpler myopic strategy and provide
the designer with a tool to gauge the expected loss with respect to
the optimal, but intractable, multi-stage policy. The bound was used
to design resource constrained, information driven sensor
management algorithms that exploit the submodularity property. 
The algorithm monotonically reduces an upper bound on the optimal solution
that permits the system designer to terminate
computation early with a near-optimal solution.  These results are
further elaborated in \cite{williams2007information,Williams:07, Williams_Fisher:07, Williams_Fisher:07AISTATS}. 

%% file: SM5_Opportunities_v9.tex
\section{Opportunities on the horizon}
\label{sec:horizon}

Despite intensive research activity over the past fifteen years, and 
particularly in the past decade, formidable challenges remain to be
addressed in order for sensor management to be genuinely viable in
large-scale sensing systems.  A central issue is computational
feasibility of even approximate methods when scaled to problems
that involve large numbers of controllable parameters, pose acute
time constraints, or can only be adequately addressed by methods that
look multiple steps ahead.

One arena of current investigation seeking to address the complexity
issue involves sparse convex optimization approaches. 
The selection of an action
sequence among a large number of possible sequences is similar to
variable selection in sparse (lasso) regression \cite{Hastie&etal:01}
and compressive sensing \cite{Candes&etal:05}, among other areas. This
insight led R.~Rangarajan {\it et al.} \cite{Rangarajan&etal:ICASSP06} to
apply convex relaxation to optimal waveform design. A similar approach was
later applied by S.~Joshi and S.~Boyd \cite{joshi2009sensor} to sensor
selection.  The use of such convex relaxation principles to develop
tractable approximations to more complex sensor management
combinatorial optimization problems, such as multi-stage planning, may
lead to computational breakthroughs.

Another circle of current research offering some promise with
regard to mitigating  complexity involves the use of statistical 
machine learning tools. Often difficult problems in one
domain can be reduced to equivalent problems in another domain for
which different and effective solution tools have been developed. For
example, the celebrated boosting method of Y.~Freund and R.~Schapire 
for learning optimal classifiers \cite{Freund&Schapire:JCSS97} was directly
motivated by optimal multi-armed bandit
strategies. Conversely, by casting offline learning of
optimal POMDP policies as an equivalent problem of learning optimal
classifiers \cite{langford2003reducing}, \cite{blatt2005nips}, 
D.~Blatt {\it et al.}
\cite{blatt2006oss} developed a boosting approach to learning optimal
sensor management policies for UXO and radar sensing applications. It
is likely that other advances in statistical machine learning can
have positive impact on sensor management.

The authors are aware of ongoing research involving new approximation
schemes that adaptively partition the information state space in an
MDP problem in a way that allows controllable tradeoff of computational
efficiency and approximation fidelity.  This work, as yet unpublished,
casts fidelity in terms of preserving the ranking of possible actions
in terms of expected loss rather than preserving the actual values of
expected loss.  

The area of adversarial sensor management, which deals with situations where
an adversary can control some aspects of the scenario to deliberately
confound the sensor manager's objectives, presents opportunities
for new sensor management research directions involving game theory and
other methods. Recent work on POMDP for smart targets,
i.e., targets that can react when they sense that they are being
probed, is a step in this direction
\cite{Kreucher&etal:DASP04},\cite{binblending09}.  Adversarial 
multi-armed bandits \cite{uchiya2010algorithms} and game
theoretic solutions to adversarial multimobile sensing
have also been proposed \cite{wei2008game}. However, there are 
presently very
few fundamental results on performance in adversarial environments;
e.g., generalizations of the non-adversarial bounds of
Casta\~{n}\'{o}n \cite{castanon2005stochastic} and  Williams
\cite{williams2007information} for POMDPs or those of 
K.~D.~Glazebrook and R.~Minty for
multi-armed bandits \cite{glazebrook2009generalized}.

%% file: SM6_Conclusion_v9.tex
\section{Concluding remarks}
\label{sec:conclusion}
In this overview article on sensor management we have described
the primary models and methods around which recent research in the
field has been centered.  We have also attempted to expose the historical
roots in classical work spanning sequential analysis, optimal design 
of experiments, information theory, and optimal control.
In our discussion of current trends and future research opportunities,
we point out formidable challenges to achieving the performance gains in
real-world systems that we believe are potentially possible. The
computational viability of scaling the methods described in this paper
to large-scale problems involving sensing systems with many 
controllable parameters, applications with fast operating tempos, and
scenarios calling for non-myopic optimization depends upon substantial
advances in efficient and certifiable approximation in all the
main components depicted in Figure \ref{fig:conceptual_block}. 

Nevertheless, there is much room for optimism.  The past two decades 
have seen intense research activity that has legitimized sensor
management as a field of study and established its mathematical 
foundations. These have drawn on, adapted, and blended ideas from
several established areas, including Markov decision processes, 
multi-armed bandit scheduling,
and information gain myopic planning.  The applications and technological
advances that spurred the profound growth of interest in sensor
management during this period continue to provide more and more opportunities
for sensor management, and in some cases demand it. In our own application
regime of surveillance and reconnaissance for security and defense applications,
future operational concepts envision increasingly versatile networked
collections of
sensor assets, a large fraction of which will be mounted on autonomous or
semi-autonomous platforms, providing situational awareness at levels of
abstraction considerably higher than target tracks and emission source localizations.
We remain hopeful that a combination of significant incremental advances
and {\it bona fide} breakthroughs will enable sensor management to rise to
meet such visions. 

In closing, we wish to acknowledge the role of numerous sponsored research programs
that have enabled and shaped the development of sensor management over the past decade.  Some such activities of which we are aware include DARPA's {\em Integrated Sensing and Processing} and {\em Waveforms for Active Sensing} programs, which ran from 2001 through 2006.  The U.S.~Department of Defense has also invested in academic research in sensor management through several Multidisciplinary University Research Initiatives (MURIs) since the early 2000s. These have been managed by DARPA, the U.S.~Air Force Office of Scientific Research (AFOSR), and the U.S.~Army Research Office (ARO). We are also aware of sponsored work through the U.S.~Air Force Research Laboratory, the Australian Defence Science and Technology Organisation, a few other government agencies, and several industrial sources. This list is by no means comprehensive, but it illustrates the recognition of sensor management as a valuable emerging area of study by major research organizations.  Further, this trend is ongoing. For example, two new MURI projects related to sensor management have recently been initiated, one by AFOSR in 2010 entitled {\em Control of Information Collection and Fusion} and the most recent by ARO in 2011 entitled {\em Value of Information for Distributed Data Fusion}.